\documentclass[draft,showpacs,showkeys,eqsecnum,nofootinbib]{revtex4}
\renewcommand{\theequation}{\arabic{equation}}
\def\beq{\begin{equation}}
\def\eeq{\end{equation}}
\def\bea{\begin{eqnarray}}
\def\eea{\end{eqnarray}}
\def\nn{\nonumber}

\begin{document}
\title{SU(3) Group Structure of Strange Flavor Hadrons}
\author{Soon-Tae Hong}
\email{soonhong@ewha.ac.kr} \affiliation{Department of Science
Education and Research Institute for Basic Sciences,\\ Ewha Womans
University, Seoul 120-750, Korea}
\date{\today}%

\begin{abstract}
We provide the isoscalar factors of the SU(3) Clebsch-Gordan
series {\bf 8}$\otimes${\bf 35} which are extensions of the
previous works of de Swart, McNamee and Chilton and play practical roles in current ongoing
strange flavor hadron physics research. To this end, we pedagogically study the SU(3) Lie algebra, its
spin symmetries, and its eigenvalues for irreducible representations. We also evaluate the values of 
the Wigner $D$ functions related to the isoscalar factors; these functions 
are immediately applicable to strange flavor hadron phenomenology.
Exploiting these SU(3) group properties associated with the spin symmetries, we 
investigate the decuplet-to-octet transition magnetic
moments and the baryon octet and decuplet magnetic moments in the flavor symmetric limit to 
construct the Coleman-Glashow-type sum rules.
\end{abstract}
\pacs{21.60.Fw, 02.20.-a, 02.20.Qs, 14.20.-c, 13.40.Gp}
\keywords{Isoscalar factor, Wigner D functions, Clebsch-Gordan series, Baryon, Strange form factor, Spin symmetries,
Coleman-Glashow sum rules}
\maketitle

\section{introduction}
\setcounter{equation}{0}
\renewcommand{\theequation}{\arabic{section}.\arabic{equation}}

Since Frisch and Stern~\cite{stern33} performed the first
measurement of the magnetic moment of the proton and obtained the
earliest experimental evidence for the internal structure of the
nucleon and Coleman and Glashow~\cite{cg} predicted the magnetic
moments of the baryon octet, there have been many interesting
developments concerning the strange flavor structures in the
nucleon and the hyperons. Recently, the experimental data for the
proton strange form factors have been made available in the
SAMPLE~\cite{sample00prl,sample00,sample04} and
HAPPEX~\cite{happex01,happex04,happex06prl,happex06plb,happex07}
Collaborations, and the predictions for the magnetic moments and
the strange form factors of the above experimental data and the other
octet baryons have been hot issues in both theoretical and experimental hadron
phenomenology~\cite{hong02,hong93,hong93oct,jenkins93plb,hong97,
musolf97,musolf98,dong98,williams99,riska00,mckeown01,musolf01,
hong03,hong04,gerasimov04,chen04,kim04,zou05,an06,xia08,wang09}.
Additionally, measurements of the baryon decuplet magnetic
moments were reported for $\mu_{\Delta^{++}}$~\cite{boss} and
$\mu_{\Omega^{-}}$~\cite{die} to yield an understanding of the hyperon
structure. The (flavor) magnetic moments and decuplet-to-octet
transition magnetic moments of the baryon decuplet have also been 
theoretically investigated in several
models~\cite{lein,sch,decup,but,jenkins94plb,jenkins94,
jenkins95,jenkins96,linde,lee1,lin2,alie00v,jenkins02prl,
kim04prd,lebed04,kim05,hong07}.

Exploiting the chiral bag model~\cite{gerry791}, we have predicted
theoretical values. Especially, after constructing the magnetic
moments of octet and decuplet baryons, we have formulated the sum
rules among the magnetic moments, which produce strange form
factor predictions
successively~\cite{hong93oct,hong93,decup,hong02,hong03,hong07}.
In the chiral theory, we need an SU(3) flavor group
analysis to construct the theoretical hadron physics formula. We
note that the SU(3) group
structure~\cite{hamermesh62,gilmore73,wybourne74,georgi99,gilmore08}
is a generic property shared by the chiral models that exploit the
hedgehog ansatz solution corresponding to the little group
SU(2)$\times Z_{2}$~\cite{jenkins94}. The SU(3) isoscalar factors are given 
in Refs. 58 and 59, which are beneficial to the
strange flavor related physics. However, in order to perform the strange hadron
physics researches involving the predictions of the ongoing
experimental data, we have necessities to update the information of
the SU(3) isoscalar factors.

In this paper, we will list these explicit values of the SU(3)
isoscalar factors for the Clebsch-Gordan (CG) series {\bf
8}$\otimes${\bf 35}, that are absent in the previous
works~\cite{deswart63,mcnamee64}, some of which are necessary
and useful in the current ongoing research. As heuristic
applications of the isoscalar factors for the series {\bf
8}$\otimes${\bf 35}, we also evaluate and summarize the values of
the Wigner $D$ functions, some of which can be directly applied
to the strange flavor hadron physics of interest. We will apply
these SU(3) group properties related with the spin symmetries 
to the baryon octet and decuplet magnetic moments and to the decuplet-to-octet transition
magnetic moments to obtain their Coleman-Glashow~\cite{cg}-type
sum rules.

This paper is organized as follows: In Section II, we will introduce
SU(3) Lie algebra and its spin symmetries. In Section III,
we will explicitly yield isoscalar factors for {\bf 8}$\otimes${\bf 35} and
investigate the Wigner $D$ functions associated
with the hadron phenomenology. The Coleman-Glashow-type sum rules will be given, 
as applications of the SU(3) group theoretical results, in Section IV. 
Section V includes a summary and discussion.

\section{SU(3) Lie algebra and spin symmetries}
\setcounter{equation}{0}
\renewcommand{\theequation}{\arabic{section}.\arabic{equation}}

In this section, we will start with the SU(3) group Lie algebra associated with
eight generators $\lambda_{a}$ $(a=1,2,...,8)$. These generators can be
expressed by using Gell-Mann matrices satisfying ${\rm tr}(\lambda_{a}\lambda_{b})=2\delta_{ab}$ and
$[\lambda_{a},\lambda_{b}]=2if_{abc}\lambda_{c}$, with
$f_{abc}=\frac{1}{4i}{\rm tr}([\lambda_{a},\lambda_{b}]\lambda_{c})$. In
hadron physics, we have $\hat{I}_{i}=\frac{1}{2}\lambda_{i}$, $(i=1,2,3)$ and
$\hat{Y}=\frac{1}{\sqrt{3}}\lambda_{8}$, which are isospin generators
and a hypercharge generator, respectiely. In particular, combining the
diagonal generators $\lambda_{3}$ and $\lambda_{8}$, one can
construct the electromagnetic charge operator $Q_{EM}$ given by the
following Gell-Mann-Nishijima relation:
\beq
Q_{EM}=e\left(\hat{I}_{3}+\frac{1}{2}\hat{Y}\right)
=\frac{e}{2}\left(\lambda_{3}+\frac{1}{\sqrt{3}}\lambda_{8}\right),
\label{qem}
\eeq
where $e$ ($e>0$) is the magnitude of the electron charge. The
other four generators $\lambda_{M}$ $(M=4,5,6,7)$ connect the
isospins and hypercharge to yield the enlarged group SU(3) from
SU(2)$\times$U(1). The finite SU(3) transformation is given as
\beq
U=e^{-i\rho\lambda_{8}/\sqrt{3}}e^{-i\alpha\lambda_{3}/2}e^{-i\beta\lambda_{2}/2}
e^{-i\gamma\lambda_{3}/2}e^{-i(\delta\lambda_{4}+\delta^{\prime}\lambda_{5}+\epsilon\lambda_{6}
+\epsilon^{\prime}\lambda_{7})}, \label{u1} \eeq which can also be 
rewritten in the form~\cite{nelson67,holland69} \beq
U=e^{-i\alpha\lambda_{3}/2}e^{-i\beta\lambda_{2}/2}
e^{-i\gamma\lambda_{3}/2}e^{-i\rho\lambda_{8}/\sqrt{3}}e^{-i\delta\lambda_{4}}e^{-i\alpha^{\prime}\lambda_{3}/2}
e^{-i\beta^{\prime}\lambda_{2}/2}
e^{-i\gamma^{\prime}\lambda_{3}/2}.\label{u2} \eeq Here, the angle
variables $\delta^{\prime}$, $\epsilon$ and $\epsilon^{\prime}$ in Eq. 
(\ref{u1}) are reshuffled to yield the new angle variables
$\alpha^{\prime}$, $\beta^{\prime}$ and $\gamma^{\prime}$ in Eq. 
(\ref{u2}), and the identity $e^{A}Be^{-A}=B+[A,B]+\frac{1}{2!}[A,[A,B]]+\cdots$ has been used.
The SU(3) group has two Casimir operators $C_{2}$ and
$C_{3}$, which are given in terms of $\lambda_{a}$ as follows:
\bea
C_{2}&=&\frac{1}{4}\sum_{a=1}^{8}\lambda_{a}^{2},\nn\\
C_{3}&=&\frac{1}{4}\lambda_{1}(\{\lambda_{4},\lambda_{6}\}+\{\lambda_{5},\lambda_{7}\})
+\frac{1}{4}\lambda_{2}(-\{\lambda_{4},\lambda_{7}\}+\{\lambda_{5},\lambda_{6}\})
+\frac{1}{4}\lambda_{3}(\lambda_{4}^{2}+\lambda_{5}^{2}-\lambda_{6}^{2}-\lambda_{7}^{2})
\nn\\
&&+\frac{1}{\sqrt{3}}\lambda_{8}\left(\frac{1}{2}(\lambda_{1}^{2}+\lambda_{2}^{2}+\lambda_{3}^{2})
-\frac{1}{6}\lambda_{8}^{2}-1-\frac{1}{4}(\lambda_{4}^{2}
+\lambda_{5}^{2}+\lambda_{6}^{2}+\lambda_{7}^{2})\right).
\eea

Next, we use $(\lambda,\mu)$ and $(Y,I,I_{3})$ to denote an irreducible
representation (IR) and a state within the IR. For instance,
$\{\lambda_{1},\lambda_{2},\lambda_{3}\}$ are the
basis states chosen such that  \beq
I^{2}=\frac{1}{4}(\lambda_{1}^{2}+\lambda_{2}^{2}+\lambda_{3}^{2}),~~~[I_{i},I_{j}]=\epsilon_{ijk}I_{k};
\eeq
then, SU(3) has the isospin rotation group SU(2) as a subgroup, as expected. In SU(3) algebra,
$\lambda_{3}$ and $\lambda_{8}$ are diagonal and satisfy \bea \langle
YII_{3}|e^{-i\alpha\lambda_{3}/2} e^{-i\beta\lambda_{2}/2}
e^{-i\gamma\lambda_{3}/2}|Y^{\prime}I^{\prime}I_{3}^{\prime}
\rangle&=&D^{I}_{I_{3}I_{3}^{\prime}}(\alpha,\beta,\gamma)
\delta_{YY^{\prime}}\delta_{II^{\prime}}\label{dsym}\\
\langle
YII_{3}|e^{-i\rho\lambda_{8}/\sqrt{3}}|Y^{\prime}I^{\prime}I_{3}^{\prime}
\rangle&=&e^{-i\rho Y^{\prime}}\delta_{YY^{\prime}}
\delta_{II^{\prime}}\delta_{I_{3}I_{3}^{\prime}}.\label{exp} \eea

Next, in order to discuss the $I$-, $U$- and $V$-spin symmetries
of the SU(3) group, we introduce the root diagram approach to the
construction of the Lie algebra of the SU(3) group which has eight
generators. Because the rank of the SU(3) group is two, one can have
the Cartan subalgebra~\cite{wybourne74,georgi99}, the
set of two commuting generators $H_{i}$ ($i=1,2$) \beq [
H_{1},~H_{2}]=0, \label{h1h2} \eeq and the other generators
$E_{\alpha}$ ($\alpha =\pm 1,\pm 2,\pm 3$) satisfying the
commutator relations \bea
\left[H_{i},~E_{\alpha}\right]&=&e^{\alpha}_{i}E_{\alpha},\nn\\
\left[E_{\alpha},~E_{\beta}\right]&=&c_{\alpha\beta}E_{\gamma},\nn\\
\left[ E_{\alpha},~E_{-\alpha}\right]&=&e^{\alpha}_{i}H_{i},
\label{commutatorshe} \eea where $e^{\alpha}_{i}$ $(i=1,2)$ is the $i$-th
component of the root vector $\hat{e}^{\alpha}$ in a two-dimensional root space 
and $c_{\alpha\beta}$ are normalization constants.
Here, $H_{i}$ is the Hermitian operator $H_{i}^{\dagger}=H_{i}$, 
and $E_{-\alpha}$ is the Hermitian conjugate of $E_{\alpha}$,
namely, $E_{\alpha}^{\dag}=E_{-\alpha}$.

\begin{figure}[h]
\setlength{\unitlength}{0.65cm}
\vskip 1.5cm
\begin{center}
\vskip-0.7cm
\begin{picture}(10,7)(-5,-4)
\thinlines \put(5.5,0){$I_{3}$}
\put(0,3.7){$Y$} \put(3,2){\circle*{0.15}}
\put(3.45,2.3){$\Delta^{++}$} \put(-3,2){\circle*{0.15}}
\put(-3.80,2.3){$\Delta^{-}$} \put(-1,2){\circle*{0.15}}
\put(-1.20,2.3){$\Delta^{0}$} \put(1,2){\circle*{0.15}}
\put(1,2.3){$\Delta^{+}$} \put(1,-2){\circle*{0.15}}
\put(1.55,-2.3){$\Xi^{*0}$} \put(-1,-2){\circle*{0.15}}
\put(-1.95,-2.3){$\Xi^{*-}$} \put(2,0){\circle*{0.15}}
\put(2.45,-0.5){$\Sigma^{*+}$} \put(0,0){\circle*{0.15}}
\put(0.45,-0.5){$\Sigma^{*0}$} \put(-2,0){\circle*{0.15}}
\put(-2.85,-0.5){$\Sigma^{*-}$} \put(0,-4){\circle*{0.15}}
\put(0.45,-4.35){$\Omega^{-}$} \put(-5,0){\vector(1,0){10.0}}
\put(0.0,-4.5){\vector(0,1){8.0}} \put(-3,2){\line(1,0){6}}
\put(-3,2){\line(1,-2){3}} \put(3,2){\line(-1,-2){3}}
\put(0.9,0.5){$\frac{1}{2}$} \put(1,-0.1){\line(0,1){0.2}}
\put(-1.5,0.5){$-\frac{1}{2}$} \put(-1,-0.1){\line(0,1){0.2}}
\put(0.3,-2.1){$-1$} \put(-0.1,-2){\line(1,0){0.2}}
\put(-2.25,2.3){$I_{+}$} \put(2.15,2.3){$I_{-}$}
\put(3.2,1){$V_{-}$} \put(-3.4,1){$U_{-}$} \put(1,-3.5){$V_{+}$}
\put(-1.2,-3.5){$U_{+}$} \thicklines \put(-3,2){\vector(1,0){2}}
\put(-3,2){\vector(1,-2){1}} \put(3,2){\vector(-1,0){2}}
\put(3,2){\vector(-1,-2){1}} \put(0,-4){\vector(1,2){1}}
\put(0,-4){\vector(-1,2){1}}
\end{picture}
\end{center}
\caption{$I_{\pm}$-, $U_{\pm}$- and $V_{\pm}$-spin symmetry
operations in the baryon decuplet.} \label{spin}
\end{figure}
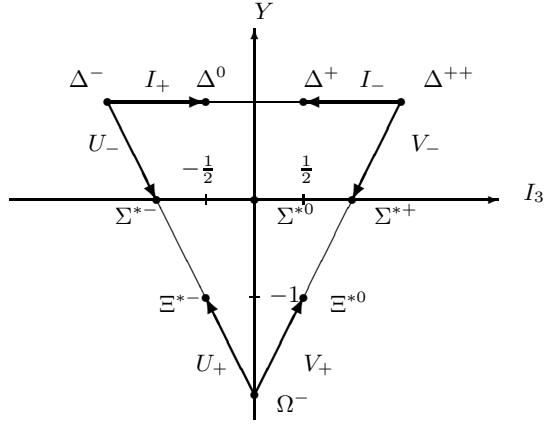

Normalizing the root vectors such that
$\sum_{\alpha}e_{i}^{\alpha}e_{j} ^{\alpha}=\delta_{ij}$, one can
choose the root vectors
\bea
\hat{e}^{1}&=&-\hat{e}^{-1}=\left(\frac{1}{\sqrt{3}},0\right),\nonumber\\
\hat{e}^{2}&=&-\hat{e}^{-2}=\left(\frac{1}{2\sqrt{3}},\frac12\right),\nonumber\\
\hat{e}^{3}&=&-\hat{e}^{-3}=\left(-\frac{1}{2\sqrt{3}},\frac12\right),
\label{rootvectors} \eea and one has
two simple roots $\hat{e}^{2}$ and $\hat{e}^{-3}$ of equal
length separated by an angle $2\pi/3$ so that one can
obtain the Dynkin diagram \cite{wybourne74,georgi99} for
the SU(3) Lie algebra given by the group theoretical symbol
$\circ\hbox{\hskip-0.5em}-\hbox{\hskip -1.0em}-\hbox{\hskip -0.5em}\circ$.
Next, the $c_{\alpha\beta}$ satisfy the identities
$c_{\alpha\beta}=-c_{\beta\alpha}=-c_{-\alpha,-\beta}=c_{\beta,-\gamma}=c_{-\gamma,\alpha}$
and $c_{\alpha\beta}c_{\alpha+\beta,\rho}+c_{\rho\alpha}
c_{\alpha+\rho,\beta}+c_{\beta\rho}c_{\beta+\rho,\alpha}=0$.
Explicitly, we have
$c_{13}=c_{3,-2}=c_{-1,2}=\frac{1}{\sqrt{6}}$~\cite{lee62}.
Moreover, the matrix representations of the SU(3) generators
$H_{i}$ and $E_{\alpha}$ can be given in terms of the Gell-Mann
matrices as \bea
H_{1}&=&\frac{1}{2\sqrt{3}}\lambda_{3},~H_{2}=\frac{1}{2\sqrt{3}}\lambda_{8},~
E_{\pm 1}=\frac{1}{2\sqrt{3}}(\lambda_{1}\pm i\lambda_{2}),\nn\\
E_{\pm 2}&=&\frac{1}{2\sqrt{3}}(\lambda_{4}\pm i\lambda_{5}),~
E_{\pm 3}=\frac{1}{2\sqrt{3}}(\lambda_{6}\pm i\lambda_{7}).
\eea

Substituting the root vectors normalized as in Eq. (\ref{rootvectors}) into 
the relations in Eqs. (\ref{h1h2}) and
(\ref{commutatorshe}), one can readily derive the commutator
relations~\cite{deswart63} \beq
\begin{array}{ll}
\left[H_{1},~H_{2}\right]=0,
&\left[H_{1},~E_{1}\right]=\displaystyle\frac{1}{\sqrt{3}}E_{1},\\
\left[H_{1},~E_{2}\right]=\displaystyle\frac{1}{2\sqrt{3}}E_{2},
&\left[H_{1},~E_{3}\right]=\displaystyle-\frac{1}{2\sqrt{3}}E_{3},\\
{\left[H_{2},~E_{1}\right]}=0,
&\left[H_{2},~E_{2}\right]=\displaystyle\frac{1}{2}E_{2},\\
\left[H_{2},~E_{3}\right]=\displaystyle\frac12 E_{3},
&\left[E_{1},~E_{-1}\right]=\displaystyle\frac{1}{\sqrt{3}}H_{1},\\
\left[E_{2},~E_{-2}\right]=\displaystyle\frac{1}{2\sqrt{3}}H_{1}+\frac12 H_{2},
&\\
\left[E_{3},~E_{-3}\right]=\displaystyle-\frac{1}{2\sqrt{3}}H_{1}+\frac12
H_{2}, &\left[E_{1},~E_{3}\right]=\displaystyle\frac{1}{\sqrt{6}}E_{2},\\
\left[E_{2},~E_{-3}\right]=\displaystyle-\frac{1}{2\sqrt{3}}H_{1}+\frac12
H_{2}, &\left[E_{1},~E_{3}\right]=\displaystyle\frac{1}{\sqrt{6}}E_{2},\\
\left[E_{2},~E_{-3}\right]=\displaystyle\frac{1}{\sqrt{6}}E_{1},
&\left[E_{-1},~E_{2}\right]=\displaystyle\frac{1}{\sqrt{6}}E_{3}.
\end{array}
\label{hhhes} \eeq

Associating the root vectors $H_{i}$ ($i=1,2$) and $E_{\alpha}$
($\alpha= \pm1,\pm 2,\pm 3$) with the physical operators $Y$,
$I_{3}$, $I_{\pm}$, $U_{\pm}$ and $V_{\pm}$ through the
definitions \beq
\begin{array}{lll}
H_{1}=\displaystyle\frac{1}{\sqrt{3}}I_{3},&H_{2}=\displaystyle\frac12 Y, &\\
E_{\pm 1}=\displaystyle\frac{1}{\sqrt{6}}I_{\pm},
&E_{\pm 2}=\displaystyle\frac{1}{\sqrt{6}}V_{\pm},
&E_{\pm 3}=\displaystyle\frac{1}{\sqrt{6}}U_{\pm},\end{array} \label{h1i3} \eeq one
can use the commutator relations in Eq. (\ref{hhhes}) to yield the
explicit expressions for the eigenvalue equations of the spin
operators in the SU(3) group~\cite{deswart63}:
\bea
I_{+}|Y,I,I_{3}\rangle&=&((I-I_{3})(I+I_{3}+1))^{\frac12}|Y,I,I_{3}+1\rangle,
\nonumber\\
I_{-}|Y,I,I_{3}\rangle&=&((I+I_{3})(I-I_{3}+1))^{\frac12}|Y,I,I_{3}-1\rangle,
\nonumber\\
U_{+}|Y,I,I_{3}\rangle&=&(a_{+}(I-I_{3}+1))^{\frac12}|Y+1,I+\frac12,
I_{3}-\frac12\rangle
-(a_{-}(I+I_{3}))^{\frac12}|Y+1,I-\frac12, I_{3}-\frac12\rangle,
\nonumber\\
U_{-}|Y,I,I_{3}\rangle&=&-(b_{+}(I+I_{3}+1))^{\frac12}|Y-1,I+\frac12,
I_{3}+\frac12\rangle
+(b_{-}(I-I_{3}))^{\frac12}|Y-1,I-\frac12, I_{3}+\frac12\rangle,
\nonumber\\
V_{+}|Y,I,I_{3}\rangle&=&(a_{+}(I+I_{3}+1))^{\frac12}|Y+1,I+\frac12,
I_{3}+\frac12\rangle
+(a_{-}(I-I_{3}))^{\frac12}|Y+1,I-\frac12, I_{3}+\frac12\rangle,\nn\\
V_{-}|Y,I,I_{3}\rangle&=&(b_{+}(I-I_{3}+1))^{\frac12}|Y-1,I+\frac12,
I_{3}-\frac12\rangle
+(b_{-}(I+I_{3}))^{\frac12}|Y-1,I-\frac12, I_{3}-\frac12\rangle.
\label{iuvplus} \eea
Figure~\ref{spin} depicts the $I_{\pm}$-, $U_{\pm}$- and
$V_{\pm}$-spin symmetry operation diagram in the case of the
decuplet baryons. In Eq. (\ref{iuvplus}), we have used the de Swart phase
convention~\cite{deswart63} and
\bea
a_{+}&=&\frac{(Y_{+}+1)
(Y_{+}+q+2)(-Y_{+}+p)}{2(I+1)(2I+1)},
\nonumber\\
a_{-}&=&\frac{Y_{-}(Y_{-}+q+1)(Y_{-}-p-1)}{2I(2I+1)},\nn\\
b_{+}&=&\frac{(Y_{-}-1)(Y_{-}+q)(Y_{-}-p-2)}{2(I+1)(2I+1)},
\nonumber\\
b_{-}&=&\frac{Y_{+}(Y_{+}+q+1)(-Y_{+}+p+1)}{2I(2I+1)},\label{applusaminus} \eea
with $Y_{\pm}=\frac12 Y\pm I+\frac{1}{3}(p-q)$. Here, $p$ and $q$
are nonnegative coefficients needed to construct bases for the IR
$D(p,q)$ of SU(3) group.  The dimension ${\bf n}$ of $D(p,q)$,
namely, the number of the basis vectors, is then given by \beq {\bf
n}=\frac{(p+1)(q+1)(p+q+2)}{2} \eeq to yield the IRs ${\bf
1}=D(0,0)$, ${\bf 3}=D(1,0)$, $\bar{\bf 3}=D(0,1)$, ${\bf
8}=D(1,1)$, ${\bf 10}=D(3,0)$, $\bar{\bf 10}=D(0,3)$, $\bar{\bf
27}=D(2,2)$, ${\bf 35}=D(4,1)$, $\bar{\bf 35}=D(1,4)$, ${\bf
28}=D(6,0)$, ${\bf 64}=D(3,3)$, $\bar{\bf 81}=D(5,2)$ and
$\bar{\bf 81}=D(2,5)$~\cite{deswart63,mcnamee64}.

\section{isoscalar factors for {\bf 8}$\otimes${\bf 35} and Wigner $D$ functions}
\setcounter{equation}{0}
\renewcommand{\theequation}{\arabic{section}.\arabic{equation}}

\begin{figure}[ht]
\setlength{\unitlength}{0.6cm}
\begin{center}
\hskip-0.7cm
\begin{picture}(3.3,4.4)(0,0)
\put(0.2,2.0){\vector(1,0){2.8}} \put(1.5,0.7){\vector(0,1){2.8}}
\put(0.3,3.8){${\bf 8}$} \put(3.2,1.9){$I_{3}$} \put(1.4,3.8){$Y$}
\put(2.1,2.0){\circle*{0.3}} \put(1.5,2.0){\circle*{0.1}}
\put(0.9,2.0){\circle*{0.1}} \put(1.8,2.6){\circle*{0.1}}
\put(1.2,2.6){\circle*{0.1}} \put(1.8,1.4){\circle*{0.1}}
\put(1.2,1.4){\circle*{0.1}} \put(1.5,2.0){\line(1,2){0.3}}
\put(1.5,2.0){\line(1,-2){0.3}} \put(2.1,2.0){\line(-1,2){0.3}}
\put(2.1,2.0){\line(-1,-2){0.3}}
\end{picture}
\hskip0.7cm
\begin{picture}(3.3,4.4)(0,0)
\put(0.2,2.0){\vector(1,0){2.8}} \put(1.5,0.7){\vector(0,1){2.8}}
\put(0.3,3.8){${\bf 10}$} \put(3.2,1.9){$I_{3}$}
\put(1.4,3.8){$Y$} \put(2.1,2.0){\circle*{0.1}}
\put(1.5,2.0){\circle*{0.1}} \put(0.9,2.0){\circle*{0.1}}
\put(1.8,2.6){\circle*{0.1}} \put(1.2,2.6){\circle*{0.1}}
\put(2.4,2.6){\circle*{0.3}} \put(0.6,2.6){\circle*{0.1}}
\put(1.8,1.4){\circle*{0.1}} \put(1.2,1.4){\circle*{0.1}}
\put(1.5,0.8){\circle*{0.1}} \put(1.5,0.8){\line(1,2){0.9}}
\end{picture}
\hskip0.7cm
\begin{picture}(3.3,4.4)(0,0)
\put(0.2,2.0){\vector(1,0){2.8}} \put(1.5,0.7){\vector(0,1){2.8}}
\put(0.3,3.8){$\bar{\bf10}$} \put(3.2,1.9){$I_{3}$}
\put(1.4,3.8){$Y$} \put(2.1,2.0){\circle*{0.1}}
\put(1.5,2.0){\circle*{0.1}} \put(0.9,2.0){\circle*{0.1}}
\put(1.5,3.2){\circle*{0.1}} \put(1.8,2.6){\circle*{0.1}}
\put(1.2,2.6){\circle*{0.1}} \put(0.6,1.4){\circle*{0.1}}
\put(1.2,1.4){\circle*{0.1}} \put(1.8,1.4){\circle*{0.1}}
\put(2.4,1.4){\circle*{0.3}} \put(2.4,1.4){\line(-1,2){0.9}}
\end{picture}
\hskip0.7cm
\begin{picture}(3.3,4.4)(0,0)
\put(0.2,2.0){\vector(1,0){3.0}} \put(1.5,0.7){\vector(0,1){2.8}}
\put(0.3,3.8){${\bf 27}$} \put(3.4,1.9){$I_{3}$}
\put(1.4,3.8){$Y$} \put(2.1,3.2){\circle*{0.1}}
\put(1.5,3.2){\circle*{0.1}} \put(0.9,3.2){\circle*{0.1}}
\put(0.6,2.6){\circle*{0.1}} \put(1.2,2.6){\circle*{0.1}}
\put(1.8,2.6){\circle*{0.1}} \put(2.4,2.6){\circle*{0.1}}
\put(0.3,2.0){\circle*{0.1}} \put(0.9,2.0){\circle*{0.1}}
\put(1.5,2.0){\circle*{0.1}} \put(2.1,2.0){\circle*{0.1}}
\put(2.7,2.0){\circle*{0.3}} \put(0.6,1.4){\circle*{0.1}}
\put(1.2,1.4){\circle*{0.1}} \put(1.8,1.4){\circle*{0.1}}
\put(2.4,1.4){\circle*{0.1}} \put(2.1,0.8){\circle*{0.1}}
\put(1.5,0.8){\circle*{0.1}} \put(0.9,0.8){\circle*{0.1}}
\put(1.5,2.0){\line(1,2){0.6}} \put(1.5,2.0){\line(1,-2){0.6}}
\put(2.7,2.0){\line(-1,2){0.6}} \put(2.7,2.0){\line(-1,-2){0.6}}
\end{picture}
\\
\vskip 1.0cm
\begin{picture}(4.5,6.2)(0,0)
\put(0.5,3.2){\vector(1,0){3.0}} \put(1.8,0.7){\vector(0,1){4.6}}
\put(0.6,5.6){${\bf 28}$} \put(3.7,3.1){$I_{3}$}
\put(1.7,5.6){$Y$} \put(3.6,4.4){\circle*{0.3}}
\put(3.0,4.4){\circle*{0.1}} \put(2.4,4.4){\circle*{0.1}}
\put(1.8,4.4){\circle*{0.1}} \put(1.2,4.4){\circle*{0.1}}
\put(0.6,4.4){\circle*{0.1}} \put(0.0,4.4){\circle*{0.1}}
\put(0.3,3.8){\circle*{0.1}} \put(0.9,3.8){\circle*{0.1}}
\put(1.5,3.8){\circle*{0.1}} \put(2.1,3.8){\circle*{0.1}}
\put(2.7,3.8){\circle*{0.1}} \put(3.3,3.8){\circle*{0.1}}
\put(3.0,3.2){\circle*{0.1}} \put(2.4,3.2){\circle*{0.1}}
\put(1.8,3.2){\circle*{0.1}} \put(1.2,3.2){\circle*{0.1}}
\put(0.6,3.2){\circle*{0.1}} \put(0.9,2.6){\circle*{0.1}}
\put(1.5,2.6){\circle*{0.1}} \put(2.1,2.6){\circle*{0.1}}
\put(2.7,2.6){\circle*{0.1}} \put(1.2,2.0){\circle*{0.1}}
\put(1.8,2.0){\circle*{0.1}} \put(2.4,2.0){\circle*{0.1}}
\put(1.5,1.4){\circle*{0.1}} \put(2.1,1.4){\circle*{0.1}}
\put(1.8,0.8){\circle*{0.1}} \put(1.8,0.8){\line(1,2){1.8}}
\end{picture}
\begin{picture}(4.5,6.2)(0,0)
\put(0.5,3.2){\vector(1,0){3.0}} \put(1.8,0.7){\vector(0,1){4.6}}
\put(0.6,5.6){${\bf 35}$} \put(3.7,3.1){$I_{3}$}
\put(1.7,5.6){$Y$} \put(3.0,4.4){\circle*{0.1}}
\put(2.4,4.4){\circle*{0.1}} \put(1.8,4.4){\circle*{0.1}}
\put(1.2,4.4){\circle*{0.1}} \put(0.6,4.4){\circle*{0.1}}
\put(0.3,3.8){\circle*{0.1}} \put(0.9,3.8){\circle*{0.1}}
\put(1.5,3.8){\circle*{0.1}} \put(2.1,3.8){\circle*{0.1}}
\put(2.7,3.8){\circle*{0.1}} \put(3.3,3.8){\circle*{0.3}}
\put(3.0,3.2){\circle*{0.1}} \put(2.4,3.2){\circle*{0.1}}
\put(1.8,3.2){\circle*{0.1}} \put(1.2,3.2){\circle*{0.1}}
\put(0.6,3.2){\circle*{0.1}} \put(0.9,2.6){\circle*{0.1}}
\put(1.5,2.6){\circle*{0.1}} \put(2.1,2.6){\circle*{0.1}}
\put(2.7,2.6){\circle*{0.1}} \put(1.2,2.0){\circle*{0.1}}
\put(1.8,2.0){\circle*{0.1}} \put(2.4,2.0){\circle*{0.1}}
\put(1.5,1.4){\circle*{0.1}} \put(2.1,1.4){\circle*{0.1}}
\put(2.1,1.4){\line(1,2){1.2}} \put(2.1,1.4){\line(-1,2){0.3}}
\put(1.8,2.0){\line(1,2){1.2}} \put(3.3,3.8){\line(-1,2){0.3}}
\end{picture}
\begin{picture}(4.5,6.2)(0,0)
\put(0.5,3.2){\vector(1,0){3.0}} \put(1.8,0.7){\vector(0,1){4.6}}
\put(0.6,5.6){$\bar{\bf 35}$} \put(3.7,3.1){$I_{3}$}
\put(1.7,5.6){$Y$} \put(1.5,5.0){\circle*{0.1}}
\put(2.1,5.0){\circle*{0.1}} \put(2.4,4.4){\circle*{0.1}}
\put(1.8,4.4){\circle*{0.1}} \put(1.2,4.4){\circle*{0.1}}
\put(0.9,3.8){\circle*{0.1}} \put(1.5,3.8){\circle*{0.1}}
\put(2.1,3.8){\circle*{0.1}} \put(2.7,3.8){\circle*{0.1}}
\put(3.0,3.2){\circle*{0.1}} \put(2.4,3.2){\circle*{0.1}}
\put(1.8,3.2){\circle*{0.1}} \put(1.2,3.2){\circle*{0.1}}
\put(0.6,3.2){\circle*{0.1}} \put(0.3,2.6){\circle*{0.1}}
\put(0.9,2.6){\circle*{0.1}} \put(1.5,2.6){\circle*{0.1}}
\put(2.1,2.6){\circle*{0.1}} \put(2.7,2.6){\circle*{0.1}}
\put(3.3,2.6){\circle*{0.3}} \put(0.6,2.0){\circle*{0.1}}
\put(1.2,2.0){\circle*{0.1}} \put(1.8,2.0){\circle*{0.1}}
\put(2.4,2.0){\circle*{0.1}} \put(3.0,2.0){\circle*{0.1}}
\put(3.0,2.0){\line(-1,2){1.2}} \put(3.0,2.0){\line(1,2){0.3}}
\put(2.1,5.0){\line(1,-2){1.2}} \put(2.1,5.0){\line(-1,-2){0.3}}
\end{picture}
\begin{picture}(4.5,6.2)(0,0)
\put(0.0,3.2){\vector(1,0){4.2}} \put(1.8,0.7){\vector(0,1){4.6}}
\put(0.6,5.6){${\bf 64}$} \put(4.4,3.1){$I_{3}$}
\put(1.7,5.6){$Y$} \put(0.9,5.0){\circle*{0.1}}
\put(1.5,5.0){\circle*{0.1}} \put(2.1,5.0){\circle*{0.1}}
\put(2.7,5.0){\circle*{0.1}} \put(3.0,4.4){\circle*{0.1}}
\put(2.4,4.4){\circle*{0.1}} \put(1.8,4.4){\circle*{0.1}}
\put(1.2,4.4){\circle*{0.1}} \put(0.6,4.4){\circle*{0.1}}
\put(0.3,3.8){\circle*{0.1}} \put(0.9,3.8){\circle*{0.1}}
\put(1.5,3.8){\circle*{0.1}} \put(2.1,3.8){\circle*{0.1}}
\put(2.7,3.8){\circle*{0.1}} \put(3.3,3.8){\circle*{0.1}}
\put(3.6,3.2){\circle*{0.3}} \put(3.0,3.2){\circle*{0.1}}
\put(2.4,3.2){\circle*{0.1}} \put(1.8,3.2){\circle*{0.1}}
\put(1.2,3.2){\circle*{0.1}} \put(0.6,3.2){\circle*{0.1}}
\put(0.0,3.2){\circle*{0.1}} \put(0.3,2.6){\circle*{0.1}}
\put(0.9,2.6){\circle*{0.1}} \put(1.5,2.6){\circle*{0.1}}
\put(2.1,2.6){\circle*{0.1}} \put(2.7,2.6){\circle*{0.1}}
\put(3.3,2.6){\circle*{0.1}} \put(0.6,2.0){\circle*{0.1}}
\put(1.2,2.0){\circle*{0.1}} \put(1.8,2.0){\circle*{0.1}}
\put(2.4,2.0){\circle*{0.1}} \put(3.0,2.0){\circle*{0.1}}
\put(0.9,1.4){\circle*{0.1}} \put(1.5,1.4){\circle*{0.1}}
\put(2.1,1.4){\circle*{0.1}} \put(2.7,1.4){\circle*{0.1}}
\put(2.7,1.4){\line(1,2){0.9}} \put(2.7,1.4){\line(-1,2){0.9}}
\put(2.7,5.0){\line(1,-2){0.9}} \put(2.7,5.0){\line(-1,-2){0.9}}
\end{picture}
\hskip0.2cm
\hskip0.2cm
\begin{picture}(4.5,6.2)(0,0)
\put(0.0,3.2){\vector(1,0){4.4}} \put(2.1,0.7){\vector(0,1){4.6}}
\put(0.6,5.6){${\bf 81}$} \put(4.6,3.1){$I_{3}$}
\put(2.0,5.6){$Y$} \put(0.6,5.0){\circle*{0.1}}
\put(1.2,5.0){\circle*{0.1}} \put(1.8,5.0){\circle*{0.1}}
\put(2.4,5.0){\circle*{0.1}} \put(3.0,5.0){\circle*{0.1}}
\put(3.6,5.0){\circle*{0.1}} \put(3.9,4.4){\circle*{0.1}}
\put(3.3,4.4){\circle*{0.1}} \put(2.7,4.4){\circle*{0.1}}
\put(2.1,4.4){\circle*{0.1}} \put(1.5,4.4){\circle*{0.1}}
\put(0.9,4.4){\circle*{0.1}} \put(0.3,4.4){\circle*{0.1}}
\put(0.0,3.8){\circle*{0.1}} \put(0.6,3.8){\circle*{0.1}}
\put(1.2,3.8){\circle*{0.1}} \put(1.8,3.8){\circle*{0.1}}
\put(2.4,3.8){\circle*{0.1}} \put(3.0,3.8){\circle*{0.1}}
\put(3.6,3.8){\circle*{0.1}} \put(4.2,3.8){\circle*{0.3}}
\put(3.9,3.2){\circle*{0.1}} \put(3.3,3.2){\circle*{0.1}}
\put(2.7,3.2){\circle*{0.1}} \put(2.1,3.2){\circle*{0.1}}
\put(1.5,3.2){\circle*{0.1}} \put(0.9,3.2){\circle*{0.1}}
\put(0.3,3.2){\circle*{0.1}} \put(0.6,2.6){\circle*{0.1}}
\put(1.2,2.6){\circle*{0.1}} \put(1.8,2.6){\circle*{0.1}}
\put(2.4,2.6){\circle*{0.1}} \put(3.0,2.6){\circle*{0.1}}
\put(3.6,2.6){\circle*{0.1}} \put(0.9,2.0){\circle*{0.1}}
\put(1.5,2.0){\circle*{0.1}} \put(2.1,2.0){\circle*{0.1}}
\put(2.7,2.0){\circle*{0.1}} \put(3.3,2.0){\circle*{0.1}}
\put(1.2,1.4){\circle*{0.1}} \put(1.8,1.4){\circle*{0.1}}
\put(2.4,1.4){\circle*{0.1}} \put(3.0,1.4){\circle*{0.1}}
\put(2.7,0.8){\circle*{0.1}} \put(2.1,0.8){\circle*{0.1}}
\put(1.5,0.8){\circle*{0.1}} \put(2.7,0.8){\line(1,2){1.5}}
\put(2.7,0.8){\line(-1,2){0.6}} \put(3.6,5.0){\line(1,-2){0.6}}
\put(3.6,5.0){\line(-1,-2){1.5}}
\end{picture}
\end{center}
\caption{Eigenvalue diagrams for the lowest irreducible representations.}
\label{eigenvalue}
\end{figure}
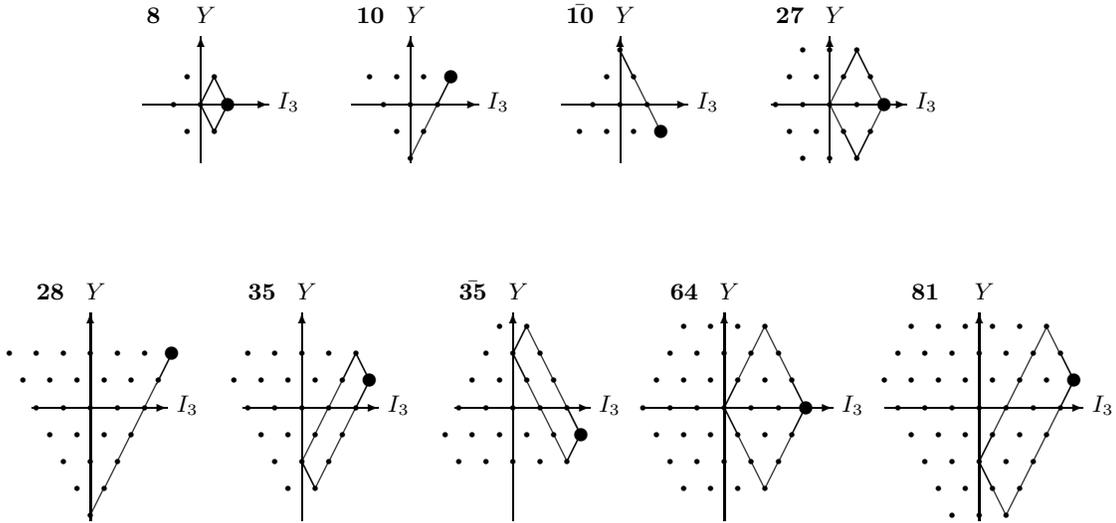

In this section, we first investigate the isoscalar factors for {\bf 8}$\otimes${\bf 35}.
To this end, we consider Fig.~\ref{eigenvalue} in which are depicted the eigenvalue diagrams for the lowest IRs.
For the dimension ${\bf n}=D(p,q)$, we have the highest eigenvalue
$e_{H}$ and its corresponding
integer hypercharge $Y_{H}$ defined as~\cite{deswart63}
\beq
e_{H}=\left(\frac{p+q}{2},\frac{p-q}{2\sqrt{3}}\right),~Y_{H}=\frac{p-q}{3}
\eeq
in $(I_{3}, Y)$ coordinates. For instances, the $e_{H}$ are denoted
by the solid disks at $\left(1,0\right)$
and $\left(\frac{3}{2},\frac{\sqrt{3}}{2}\right)$ in the diagrams in Fig.~\ref{eigenvalue} for the ${\bf 8}$ and
the ${\bf 10}$, respectively.

Starting from the solid disk $e_{H}$ for a given dimension and applying to the solid disk
the spin operators $U_{\pm}$ and
$V_{\pm}$ and the relations in Eq. (\ref{iuvplus}), we construct effectively the IRs $(I_{3},~Y)$
denoted by the points along the lines indicated in Fig.~\ref{eigenvalue}.
Similarly, we act the spin operator $I_{\pm}$ on the solid disks and points
and use the relations in Eq. (\ref{iuvplus}) to yield
the remnant IRs $(I_{3},~Y)$ denoted by the points in Fig.~\ref{eigenvalue} so that
we can derive the isoscalar
factors of the SU(3) group for the CG series, as shown in Table 1. The CG
coefficients of SU(3) group are given by~\cite{deswart63,chen02}
\beq
\left(
\begin{array}{lll}
\mu_{1} &\mu_{2} &\mu_{\gamma}\\
\nu_{1} &\nu_{2} &\nu
\end{array}
\right)=C_{I_{1}I_{2}I}^{I_{1z}I_{2z}I_{z}}
\left(
\begin{array}{ll}
\mu_{1} &\mu_{2}\\
Y_{1}I_{1} & Y_{2}I_{2}
\end{array}\Big|
\begin{array}{l}
\mu_{\gamma}\\
YI
\end{array}
\right),
\eeq
where the first part of the right-hand side is the CG coefficient of the SU(2) group
and the second one is the isoscalar factor of the SU(3) group. If the SU(3) CG coefficients are 
to be uniquely evaluated, it suffices to give the SU(3) isoscalar factors, because 
the SU(2) CG coefficients are well known. In Table 1, we list the isoscalar factors
of the SU(3) group for the CG series {\bf 8}$\otimes${\bf 35} with $\mu_{1}={\bf 8}$
and $\mu_{2}={\bf 35}$. In the first row of each table in Table 1, we have $(Y, I)$ for $\mu_{\gamma}$ being
given by the right-hand side of the CG series 
{\bf 8}$\otimes${\bf 35}={\bf 81}$\oplus${\bf 64}$\oplus${\bf 35}$\oplus${\bf 35}$\oplus${\bf
28}$\oplus${\bf 27}$\oplus${\bf 10}.  In the following rows, we have two pairs for $(Y_{1},I_{1})$ of {\bf 8}
and $(Y_{2},I_{2})$ of {\bf 35}, and the corresponding SU(3) isoscalar factor values
under the dimensions $\mu_{\gamma}$. The global signs in Table 1 are fixed to be
consistent with those in the previous works~\cite{deswart63,mcnamee64} by ensuring the
fact that each submatrix is unitary.

Finally, exploiting the isoscalar factors obtained in Table 1, we
evaluate in Table 2 the explicit expectation values of Wigner $D$
functions such as
\bea
D_{33}^{8}&=&\langle 010|D^{8}|010\rangle,~
D_{38}^{8}=\langle 010|D^{8}|000\rangle,\nn\\
D_{83}^{8}&=&\langle 000|D^{8}|010\rangle,~ D_{88}^{8}=\langle 000|D^{8}|000\rangle,\nn\\
D_{33}^{10}&=&\langle 010|D^{10}|010\rangle,~ D_{33}^{\bar{10}}=\langle 010|D^{\bar{10}}|010\rangle,\nn\\
D_{33}^{27}&=&\langle 010|D^{27}|010\rangle,~ D_{83}^{27}=\langle
000|D^{27}|010\rangle. \label{dddd} \eea In the SU(3) strange
hadron physics, the expectation value of $D_{ab}$ in the
transition $B_{1}\rightarrow B_{2}$ is given by \beq
\langle\lambda_{2} B_{2}|D_{ab}|\lambda_{1} B_{1}\rangle= \int
dA~\Phi_{B_{2}}^{\lambda_{2}*}
D_{ab}(A)\Phi_{B_{1}}^{\lambda_{1}},\label{b2b1} \eeq where \beq
D_{ab}(A)= \frac{1}{3}{\rm tr}~(A^{\dag}\lambda_{a}A\lambda_{b}).
\eeq Here, one notes that the wavefunction $\Phi_{B}^{\lambda}$ for
the baryon $B$ with quantum numbers $(\alpha)=(Y,I,I_{3})$ and
$(\beta)=(Y_{R},S,-S_{3})$ are given in IRs $\lambda$ by 
\beq
\Phi^{\lambda}_{(\alpha)(\beta)}(A)=\sqrt{\lambda}\langle
Y,I,I_{3}|D^{\lambda}(A)|Y_{R},S,-S_{3}\rangle,
\label{bwf}
\eeq 
where $Y$,
$I$ and $S$ are the hypercharge, isospin and spin of the hyperon
$B$, and the right hypercharge $Y_{R}$ is given by
$Y_{R}=\frac{1}{3}N_{c}$ due to the Wess-Zumino constraint to
yield $Y_{R}=1$ for the $N_{c}=3$ case. Next, we have 
\beq 
\int
dA~\Phi_{(\alpha_{2})(\beta_{2})}^{\lambda_{2}*}(A) \langle
\alpha|D^{\lambda}(A)|\beta\rangle
\Phi_{(\alpha_{1})(\beta_{1})}^{\lambda_{1}}(A)
=\sqrt{\frac{\lambda_{1}}{\lambda_{2}}}~\sum_{\gamma}\left(
\begin{array}{lll}
\lambda_{1} &\lambda &\lambda_{2\gamma}\\
\alpha_{1} &\alpha &\alpha_{2}
\end{array}
\right) \left(
\begin{array}{lll}
\lambda_{1} &\lambda &\lambda_{2\gamma}\\
\beta_{1} &\beta &\beta_{2}
\end{array}
\right), \eeq where the summation runs over the independent IRs in
the process $\lambda_{1}\otimes\lambda\rightarrow \lambda_{2}$.

Because the coefficients in the sum rules for the baryon magnetic moments
and form factors are solely given by the SU(3) group
structure of the chiral models, these Wigner $D$ functions can be practically used to 
the strange flavor hadron phenomenology research using the hedgehog ansatz solution
corresponding to the little group SU(2)$\times Z_{2}$. In this kind of task, it is also powerful to
use the mathematical theorem that the tensor product of the Wigner $D$ functions can be decomposed
into sum of the single $D$ functions~\cite{deswart63},
\beq
D_{\nu_{1}\lambda_{1}}^{\mu_{1}}D_{\nu_{2}\lambda_{2}}^{\mu_{2}}=\sum_{\mu\nu\lambda\gamma}
\left(
\begin{array}{lll}
\mu_{1} &\mu_{2} &\mu_{\gamma}\\
\nu_{1} &\nu_{2} &\nu
\end{array}
\right)
\left(\begin{array}{lll}
\mu_{1} &\mu_{2} &\mu_{\gamma}\\
\lambda_{1} &\lambda_{2} &\lambda
\end{array}
\right)
D_{\nu\lambda}^{\mu}.
\eeq

\section{Coleman-Glashow-type sum rules}
\setcounter{equation}{0}
\renewcommand{\theequation}{\arabic{section}.\arabic{equation}}

As applications of the above SU(3) group theoretical properties
associated with the spin symmetries of our interest, we
investigate Coleman-Glashow-type sum rules in the SU(3) flavor
symmetric limit with the chiral symmetry-breaking masses
$m_{u}=m_{d}=m_{s}$, $m_{K}=m_{\pi}$ and the decay constants
$f_{K}=f_{\pi}$. To this end in the limit, for instance, we
introduce the topological Skyrmion
model~\cite{skyrme61,anw83,hong02}, which is one of the chiral
models used in nuclear phenomenology. The Skyrmion soliton
Lagrangian with the SU(3) flavor group is given by a nonlinear equation 
of the form
\beq {\cal
L}=-\frac{1}{4}f_{\pi}^{2}{\rm
tr}(l_{\mu}l^{\mu})+\frac{1}{32e^{2}}{\rm
tr}[l_{\mu},l_{\nu}]^{2}+{\cal L}_{WZW}, \eeq where $f_{\pi}$ and
$e$ are the pion decay constant and the Skyrmion parameter and
$l_{\mu}=U^{\dagger}\partial _{\mu}U$. The chiral field
$U=e^{i\lambda_{a}\pi_{a}/f_{\pi}} \in$ SU(3) is described by the
pseudoscalar meson fields $\pi_{a}$ $(a=1,...,8)$ and the Gell-Mann
matrices $\lambda_{a}$, with $\lambda_{a}\lambda_{b}=\frac{2}{3}
\delta_{ab}+(if_{abc}+d_{abc})\lambda_{c}$. The WZW term ${\cal L}_{WZW}$ 
is described by the action~\cite{witten83wzw1,witten83wzw2} 
\beq
\Gamma_{WZW}=-\frac{iN_{c}}{240\pi^{2}}\int_{{\sf M}}d^{5}x~\epsilon^{\mu\nu\alpha\beta\gamma}{\rm
tr}(l_{\mu}l_{\nu}l_{\alpha}l_{\beta} l_{\gamma}), \label{wzwterm}
\eeq where $N_{c}$ is the number of colors and the integral is
done on the five-dimensional manifold ${\sf M}=V_{3}\times
S^{1}\times I$, with the three-space volume $V_{3}$, the
compactified time $S^{1}$ and the unit interval $I$ needed for a
local form of WZW term.

The Noether theorem then yields the flavor octet vector currents
(FOVCs) $J_{V}^{\mu a}$ $(a=1,...,8)$ from the derivative terms in
the above Skyrmion Lagrangian as follows: 
\bea J_{V}^{\mu
a}&=&-\frac{i}{2}f_{\pi}^{2}{\rm
tr}\left(\frac{\lambda_{a}}{2}l^{\mu}+(U\leftrightarrow
U^{\dagger})\right) +\frac{i}{8e^2}{\rm
tr}\left(\left[\frac{\lambda_{a}}{2},l^{\nu}\right][l^{\mu},l^{\nu}]+(U\leftrightarrow
U^{\dagger})\right)\nonumber\\
& &+\frac{N_{c}}{48\pi^{2}}\epsilon^{\mu\nu\alpha\beta}{\rm
tr}\left( \frac{\lambda_{a}}{2}l_{\nu}l_{\alpha}l_{\beta}-
(U\leftrightarrow U^{\dagger})\right), \label{jvmua} \eea with
$\epsilon^{0123}=1$. Exploiting the above FOVCs, we next calculate
the EM currents $J_{EM}^{\mu}$ as follows: \beq
J_{EM}^{\mu}=J_{V}^{\mu 3}+\frac{1}{\sqrt{3}}J_{V}^{\mu 8}, \eeq
from which we can construct the magnetic moment operators defined
by \beq \hat{\mu}^{i}=\frac{1}{2}\int
d^{3}x~\epsilon^{ijk}x^{j}J_{EM}^{k}.\label{muiop} \eeq 
For given operators, we can evaluate the matrix elements of the form factors or the transition magnetic moments 
for the diagonal or the off-diagonal case, respectively. For instance, with the spinning chiral
model ansatz in the SU(3) chiral models, the magnetic
moment operators in Eq. (\ref{muiop}) take the following form
\beq
\hat{\mu}^{i}=\hat{\mu}^{i(3)}+\frac{1}{\sqrt{3}}\hat{\mu}^{i(8)}.\label{mu}
\eeq
Here, the $\hat{\mu}^{i(a)}$ $(a=1,2,\cdots,8)$ are given by
\beq
\hat{\mu}^{i(a)}=\frac{N_{c}}{2\sqrt{3}}{\cal M}D_{a8}^{8}\hat{J}_{i}-{\cal N}D_{ai}^{8}+\cdots,
\label{muia2}
\eeq
where $\hat{J}_{i}=-\hat{T}_{i}^{R}$ are the SU(2) spin operators, with $\hat{T}_{i}^{R}$ being the right SU(3)
isospin operators along the isospin direction, and the inertia parameters ${\cal M}$ and ${\cal N}$
depend on the properties of the given SU(3) chiral model.
Here, the ellipsis stands for other contributions to the baryon magnetic moments $\mu_{B}$ of the baryon $B$,
for instance, contributions in addition to those of the chiral symmetric limit~\cite{hong02}. In the Yabu-Ando scheme~\cite{yabu88},
we also need some additional terms in $\mu_{B}$. The Wigner $D$ functions in the above
operators in Eq. (\ref{muia2}) can be used in evaluating their matrix elements or expectation elements of the form factors or the transition magnetic moments via Eq. (\ref{b2b1}). 

Specifically, exploiting this operator in Eq. (\ref{mu}), together with the baryon wave
function in Eq. (\ref{bwf}), we can evaluate the decuplet-to-octet
transition magnetic moments for ${\bf
10}(S_{3}=1/2)\rightarrow{\bf 8} (S_{3}=1/2)+\gamma$ to yield the
$V$-spin symmetry sum rules \beq
\mu_{\Sigma^{+}\Sigma^{*+}}=\mu_{\Xi^{0}\Xi^{*0}}, \eeq the
$U$-spin symmetry ones \beq
\mu_{p\Delta^{+}}=-\mu_{\Sigma^{+}\Sigma^{*+}},~~~
\mu_{\Sigma^{-}\Sigma^{*-}}=\mu_{\Xi^{-}\Xi^{*-}}, \eeq the
$I$-spin symmetry ones \beq
\mu_{p\Delta^{+}}=\mu_{n\Delta^{0}},~~~2\mu_{\Sigma^{0}\Sigma^{*0}}=\mu_{\Sigma^{+}\Sigma^{*+}}+\mu_{\Sigma^{-}\Sigma^{*-}},
\eeq and the other ones \beq
\mu_{\Sigma^{+}\Sigma^{*+}}+\mu_{\Sigma^{-}\Sigma^{*-}}=\mu_{\Xi^{0}\Xi^{*0}}+\mu_{\Xi^{-}\Xi^{*-}},~~~
\mu_{\Sigma^{0}\Sigma^{*0}}=-\sqrt{3}\mu_{\Lambda\Sigma^{*0}}.
\eeq
In the strange flavor channel of the decuplet-to-octet transition magnetic moments, we construct
the $s$-flavor currents $J^{\mu(s)}_{EM}$ by replacing the EM charge operator $Q_{EM}$ in Eq. 
(\ref{qem}) with the $s$-flavor EM charge operator $Q_{s}$. Here, one notes that by defining the
$q$-flavor projection operators
\bea
\hat{P}_{u}&=&\frac{1}{3}+\frac{1}{2}\lambda_{3}+\frac{1}{2\sqrt{3}}\lambda_{8},\nn\\
\hat{P}_{d}&=&\frac{1}{3}-\frac{1}{2}\lambda_{3}+\frac{1}{2\sqrt{3}}\lambda_{8},\nn\\
\hat{P}_{s}&=&\frac{1}{3}-\frac{1}{2\sqrt{3}}\lambda_{8},
\eea
satisfying $\hat{P}_{q}^{2}=\hat{P}_{q}$ and $\sum_{q}\hat{P}_{q}=1$, we can readily obtain the
$q$-flavor EM charge operators $Q_{q}=Q_{EM}\hat{P}_{q}=Q_{q}\hat{P}_{q}$. The EM currents are then
split into three pieces, $J^{\mu}_{EM}=J^{\mu(u)}_{EM}+J^{\mu(d)}_{EM}+J^{\mu(s)}_{EM}$. Exploiting
the $s$-flavor EM currents $J^{\mu(s)}_{EM}$ in the SU(3) flavor symmetric limit, we find the symmetry identities
\beq
\mu_{N\Delta}^{(s)}=\mu_{\Lambda\Sigma^{*0}}^{(s)},~~~\mu_{\Sigma\Sigma^{*}}^{(s)}=\mu_{\Xi\Xi^{*}}^{(s)}
\eeq
and their sum rules
\beq
\mu_{N\Delta}^{(s)}+\mu_{\Sigma\Sigma^{*}}^{(s)}=\mu_{\Lambda\Sigma^{*0}}^{(s)}+\mu_{\Xi\Xi^{*}}^{(s)}.
\eeq

Next, we construct the octet magnetic moments to yield the
$V$-spin symmetry sum rule \beq
\mu_{p}+\mu_{\Sigma_{-}}=-2\mu_{\Lambda}, \eeq the $U$-spin
symmetry ones \beq
\mu_{\Sigma^{+}}=\mu_{p},~~~\mu_{\Xi^{0}}=\mu_{n},~~~\mu_{\Xi^{-}}=\mu_{\Sigma^{-}},\label{cgu}
\eeq and the $I$-spin symmetry ones
\beq 2\mu_{\Sigma^{0}}=\mu_{\Sigma^{+}}+\mu_{\Sigma^{-}}.\label{cgi} \eeq
Finally, exploiting the decuplet baryon magnetic moments, we find
the $V$-spin symmetry sum rules \beq
\mu_{\Delta^{+}}+\mu_{\Sigma^{*0}}+\mu_{\Xi^{*-}}=0,~~~
\mu_{\Sigma^{*+}}+\mu_{\Xi^{*0}}+\mu_{\Omega^{-}}=0
\eeq
and their other sum rules
\beq
\mu_{\Delta^{++}}+\mu_{\Xi^{*0}}+\mu_{\Omega^{-}}-\mu_{\Sigma^{*+}}=0,~~~
\mu_{\Delta^{++}}+2\mu_{\Omega^{-}}=0.
\eeq
We also obtain the $U$-spin symmetry sum rules \beq
\mu_{\Delta^{-}}=\mu_{\Sigma^{*-}}=\mu_{\Xi^{*-}}=\mu_{\Omega^{-}},~~~
\mu_{\Delta^{0}}=\mu_{\Sigma^{*0}}=\mu_{\Xi^{*0}},~~~\mu_{\Delta^{+}}=\mu_{\Sigma^{+}}\label{cgu2}
\eeq and the $I$-spin symmetry ones \beq
2\mu_{\Sigma^{*0}}=\mu_{\Sigma^{*+}}+\mu_{\Sigma^{*-}},~~~
\mu_{\Delta^{-}}+\mu_{\Delta^{++}}=\mu_{\Delta^{0}}+\mu_{\Delta^{+}}.\label{cgi2}
\eeq
Here, we have included Eqs. (\ref{cgu}) and (\ref{cgi}) in Refs. 2 and 12,
and Eqs. (\ref{cgu2}) and (\ref{cgi2}) in  Ref. 36 for the sake of
completeness.

\section{Conclusions}
\setcounter{equation}{0}
\renewcommand{\theequation}{\arabic{section}.\arabic{equation}}

In summary, we have used some group theoretical formulae to
produce the isoscalar factors of the SU(3) Clebsch-Gordan series
{\bf 8}$\otimes${\bf 35}, which play a central role in the strange
flavor hadron phenomenology. This is an extension of the work of de
Swart~\cite{deswart63} in which the isoscalar factors of the SU(3)
Clebsch-Gordan series {\bf 8}$\otimes${\bf 8}, {\bf
8}$\otimes${\bf 10}, {\bf 8}$\otimes${\bf 27}, {\bf
10}$\otimes${\bf 10} and {\bf 10}$\otimes$$\bar{\bf 10}$ are
listed. Moreover, we also computed the values of the Wigner $D$
functions which are related to these isoscalar factors and are 
directly usable in the soliton model calculations. We also
described the details of how to obtain the isoscalar factors by
exploiting the eigenvalue diagrams for the lowest irreducible
representations in the SU(3) group theory.

Now, we have a couple of points to address. In the SU(3) chiral models, we assume maximal
symmetry to describe the hedgehog solution $U_{0}$ embedded in the SU(2) isospin subgroup 
of SU(3) as follows:
\beq
U_{0}=\left(\begin{array}{cc}
e^{i\vec{\tau}\cdot\hat{r}\theta(r)} &0\\
0                                    &1
\end{array}\right),
\eeq where $\tau_{i} (i=1,2,3)$ are the Pauli matrices,
$\hat{r}=\vec{x}/r$ and $\theta(r)$ is the chiral angle determined
by minimizing the static mass of the baryon described by using the
chiral models. We now emphasize that the SU(3) group structure in
the above sum rules is a generic property shared by chiral
models that use the hedgehog ansatz solution corresponding to the
little group SU(2)$\times{\bf
Z}_{2}$~\cite{jenkins94,jenkins95,jenkins96}. In the chiral
perturbation theory to which the hedgehog ansatz does not apply,
one cannot, thus, obtain the above sum rules, even though the SU(3)
flavor group is used in the theory. Here, one notes that the
quantum numbers such as spin, isospin and hypercharge in Eq. 
(\ref{bwf}) can be obtained in the chiral models by quantizing the
zero modes associated with the slow collective rotation \beq
U_{0}\rightarrow AU_{0}A^{\dagger} \eeq on the SU(3) group
manifold where $A(t)\in$ SU(3) is the time-dependent collective
variable restrained by the WZW constraint.

\acknowledgments The author is grateful to R. D. McKeown for the hospitality of the Kellogg Radiation
Laboratory at Caltech where a part of this work has been done.

\begin{table*}
\caption{Isoscalar factors for {\bf 8}$\otimes${\bf 35}.
In the first row of each table, we have $(Y, I)$ for $\mu_{\gamma}$ being
given by the right-hand side of the Clebsch-Gordon series {\bf 8}$\otimes${\bf 35}={\bf
81}$\oplus${\bf 64}$\oplus${\bf 35}$\oplus${\bf 35}$\oplus${\bf
28}$\oplus${\bf 27}$\oplus${\bf 10}.  In the following rows, two pairs for $(Y_{1},I_{1})$ of {\bf 8}
and $(Y_{2},I_{2})$ of {\bf 35} are given together with the corresponding SU(3) isoscalar factor values
under the dimensions $\mu_{\gamma}$.} \vskip 0.7cm
\begin{center}
\begin{tabular}{rrr}
\hline
$Y=3$ & $I=\frac{5}{2}$ & ${\bf 81}$ \\
\hline
1, $\frac{1}{2}$ & 2,2 & 1 \\
\hline
\end{tabular}
\hskip 1cm
\begin{tabular}{rrr}
\hline
$Y=3$ & $I=\frac{3}{2}$ & ${\bf 64}$ \\
\hline
1, $\frac{1}{2}$ & 2, 2 & $-1$ \\
\hline
\end{tabular}
\hskip 1cm
\begin{tabular}{rrrr}
\hline
$Y=2$ & $I=3$ & ${\bf 81}$ & {\bf 28} \\
\hline
1, $\frac{1}{2}$ & 1, $\frac{5}{2}$ & $\sqrt{\frac{1}{2}}$ & $-\sqrt{\frac{1}{2}}$ \\
0, 1 & 2, 2 & $\sqrt{\frac{1}{2}}$ & $\sqrt{\frac{1}{2}}$ \\
\hline
\end{tabular}
\vskip 0.7cm
\begin{tabular}{rrrrrr}
\hline
$Y=2$ & $I=2$ & ${\bf 81}$ & ${\bf 64}$ & ${\bf 35}_{S}$ & ${\bf 35}_{A}$ \\
\hline
1, $\frac{1}{2}$ & 1, $\frac{5}{2}$ & $\sqrt{\frac{1}{200}}$ & $-\sqrt{\frac{8}{25}}$
& $\sqrt{\frac{81}{180}}$ & $\sqrt{\frac{81}{360}}$\\
1, $\frac{1}{2}$ & 1, $\frac{5}{2}$ & $\sqrt{\frac{144}{200}}$ & $\sqrt{\frac{2}{25}}$
& $-\sqrt{\frac{4}{180}}$ & $\sqrt{\frac{64}{360}}$\\
0, 1 & 2, 2 & $-\sqrt{\frac{10}{200}}$ & $-\sqrt{\frac{5}{25}}$ & $-\sqrt{\frac{90}{180}}$
& $\sqrt{\frac{90}{360}}$\\
0, 0 & 2, 2 & $\sqrt{\frac{45}{200}}$ & $-\sqrt{\frac{10}{25}}$ & $-\sqrt{\frac{5}{180}}$
& $-\sqrt{\frac{125}{360}}$\\
\hline
\end{tabular}
\hskip 1cm
\begin{tabular}{rrrr}
\hline
$Y=2$ & $I=1$ & ${\bf 64}$ & {\bf 27} \\
\hline
1, $\frac{1}{2}$ & 1, $\frac{3}{2}$ & $-\sqrt{\frac{6}{7}}$ & $-\sqrt{\frac{1}{7}}$ \\
0, 1 & 2, 2 & $\sqrt{\frac{1}{7}}$ & $-\sqrt{\frac{6}{7}}$ \\
\hline
\end{tabular}
\vskip 0.7cm
\begin{tabular}{rrr}
\hline
$Y=1$ & $I=\frac{7}{2}$ & ${\bf 81}$ \\
\hline
0, 1 & 1, $\frac{5}{2}$ & $1$ \\
\hline
\end{tabular}
\hskip 1cm
\begin{tabular}{rrrrrrr}
\hline
$Y=1$ & $I=\frac{5}{2}$ & ${\bf 81}$ & {\bf 64} & ${\bf 35}_{S}$ & ${\bf 35}_{A}$ & {\bf 28}\\
\hline
1, $\frac{1}{2}$ & 0, 2 & $\sqrt{\frac{280}{800}}$ & $\sqrt{\frac{5}{75}}$
& $-\sqrt{\frac{120}{720}}$ & $\sqrt{\frac{120}{1440}}$
& $-\sqrt{\frac{40}{120}}$ \\
0, 1 & 1, $\frac{5}{2}$ & $-\sqrt{\frac{9}{800}}$ & $-\sqrt{\frac{14}{75}}$
& $-\sqrt{\frac{21}{720}}$ & $\sqrt{\frac{1029}{1440}}$
& $\sqrt{\frac{7}{120}}$ \\
0, 1 & 1, $\frac{3}{2}$ & $\sqrt{\frac{336}{800}}$ & $\sqrt{\frac{6}{75}}$
& $\sqrt{\frac{64}{720}}$ & $\sqrt{\frac{16}{1440}}$
& $\sqrt{\frac{48}{120}}$ \\
0, 0 & 1, $\frac{5}{2}$ & $\sqrt{\frac{105}{800}}$ & $-\sqrt{\frac{30}{75}}$
& $\sqrt{\frac{245}{720}}$ & $-\sqrt{\frac{5}{1440}}$
& $-\sqrt{\frac{15}{120}}$ \\
$-1$, $\frac{1}{2}$ & 2, 2 & $\sqrt{\frac{70}{800}}$ & $-\sqrt{\frac{20}{75}}$
& $-\sqrt{\frac{270}{720}}$ & $-\sqrt{\frac{270}{1440}}$
& $\sqrt{\frac{10}{120}}$ \\
\hline
\end{tabular}
\vskip 0.7cm
\begin{tabular}{rrrrrrrr}
\hline
$Y=1$ & $I=\frac{3}{2}$ & ${\bf 81}$ & {\bf 64} & ${\bf 35}_{S}$ & ${\bf 35}_{A}$ & {\bf 27}
& {\bf 10} \\
\hline
1, $\frac{1}{2}$ & 0, 2 & $\sqrt{\frac{15}{1400}}$ & $-\sqrt{\frac{45}{175}}$
& $\sqrt{\frac{245}{720}}$ & $\sqrt{\frac{80}{360}}$
& $-\sqrt{\frac{45}{560}}$ & $\sqrt{\frac{5}{56}}$ \\
1, $\frac{1}{2}$ & 0, 1 & $\sqrt{\frac{675}{1400}}$ & $\sqrt{\frac{25}{175}}$
& $-\sqrt{\frac{25}{720}}$ & $\sqrt{\frac{100}{360}}$
& $\sqrt{\frac{25}{560}}$ & $\sqrt{\frac{1}{56}}$ \\
0, 1 & 1, $\frac{5}{2}$ & $-\sqrt{\frac{2}{1400}}$ & $\sqrt{\frac{6}{175}}$
& $-\sqrt{\frac{96}{720}}$ & $-\sqrt{\frac{6}{360}}$
& $-\sqrt{\frac{216}{560}}$ & $\sqrt{\frac{24}{56}}$ \\
0, 1 & 1, $\frac{3}{2}$ & $-\sqrt{\frac{108}{1400}}$ & $-\sqrt{\frac{49}{175}}$
& $-\sqrt{\frac{289}{720}}$ & $\sqrt{\frac{49}{360}}$
& $\sqrt{\frac{49}{560}}$ & $\sqrt{\frac{1}{56}}$ \\
0, 0 & 1, $\frac{3}{2}$ & $\sqrt{\frac{540}{1400}}$ & $-\sqrt{\frac{45}{175}}$
& $-\sqrt{\frac{45}{720}}$ & $-\sqrt{\frac{45}{360}}$
& $-\sqrt{\frac{45}{560}}$ & $-\sqrt{\frac{5}{56}}$ \\
$-1$, $\frac{1}{2}$ & 2, 2 & $-\sqrt{\frac{60}{1400}}$ & $\sqrt{\frac{5}{175}}$
& $-\sqrt{\frac{20}{720}}$ & $\sqrt{\frac{80}{360}}$
& $-\sqrt{\frac{180}{560}}$ & $-\sqrt{\frac{20}{56}}$ \\
\hline
\end{tabular}
\hskip 1cm
\begin{tabular}{rrrr}
\hline
$Y=1$ & $I=\frac{1}{2}$ & {\bf 64} & {\bf 27} \\
\hline
1, $\frac{1}{2}$ & 0, 1 & $-\sqrt{\frac{5}{7}}$ & $-\sqrt{\frac{2}{7}}$ \\
0, 1 & 1, $\frac{3}{2}$ & $\sqrt{\frac{2}{7}}$ & $-\sqrt{\frac{5}{7}}$ \\
\hline
\end{tabular}
\vskip 0.7cm
\begin{tabular}{rrrr}
\hline
$Y=0$ & $I=3$ & ${\bf 81}$ & ${\bf 64}$ \\
\hline
0, 1 & 0, 2 & $\sqrt{\frac{4}{5}}$ & $\sqrt{\frac{1}{5}}$ \\
$-1$, $\frac{1}{2}$ & 1, $\frac{5}{2}$ & $\sqrt{\frac{1}{5}}$ & $-\sqrt{\frac{4}{5}}$ \\
\hline
\end{tabular}
\hskip 1cm
\begin{tabular}{rrrrrrrr}
\hline
$Y=0$ & $I=2$ & ${\bf 81}$ & ${\bf 64}$ & ${\bf 35}_{S}$ & ${\bf 35}_{A}$ & {\bf 28} & {\bf 27} \\
\hline
1, $\frac{1}{2}$ & $-1$, $\frac{3}{2}$ & $\sqrt{\frac{90}{400}}$ & $\sqrt{\frac{20}{175}}$
& $-\sqrt{\frac{45}{180}}$
& $\sqrt{\frac{90}{720}}$ & $-\sqrt{\frac{10}{50}}$ & $\sqrt{\frac{120}{1400}}$ \\
0, 1 & 0, 2 & $-\sqrt{\frac{5}{400}}$ & $-\sqrt{\frac{40}{175}}$ & 0 & $\sqrt{\frac{405}{720}}$
& $\sqrt{\frac{5}{50}}$ & $\sqrt{\frac{135}{1400}}$ \\
0, 1 & 0, 1 & $\sqrt{\frac{135}{400}}$ & $\sqrt{\frac{30}{175}}$ & $-\sqrt{\frac{30}{180}}$
& $\sqrt{\frac{15}{720}}$
& $\sqrt{\frac{15}{50}}$ & $-\sqrt{\frac{5}{1400}}$ \\
0, 0 & 0, 2 & $\sqrt{\frac{90}{400}}$ & $-\sqrt{\frac{45}{175}}$ & $\sqrt{\frac{20}{180}}$
& $\sqrt{\frac{10}{720}}$
& $-\sqrt{\frac{10}{50}}$ & $-\sqrt{\frac{270}{1400}}$ \\
$-1$, $\frac{1}{2}$ & 1, $\frac{5}{2}$ & $-\sqrt{\frac{8}{400}}$ & $\sqrt{\frac{4}{175}}$
& $-\sqrt{\frac{36}{180}}$
& $\sqrt{\frac{72}{720}}$ & $\sqrt{\frac{2}{50}}$ & $-\sqrt{\frac{864}{1400}}$ \\
$-1$, $\frac{1}{2}$ & 1, $\frac{3}{2}$ & $\sqrt{\frac{72}{400}}$ & $-\sqrt{\frac{36}{175}}$
& $-\sqrt{\frac{49}{180}}$
& $-\sqrt{\frac{128}{720}}$ & $\sqrt{\frac{8}{50}}$ & $\sqrt{\frac{6}{1400}}$ \\
\hline
\end{tabular}
\vskip 0.7cm
\begin{tabular}{rrrrrrrr}
\hline
$Y=0$ & $I=1$ & {\bf 81} & {\bf 64} & ${\bf 35}_{S}$ & ${\bf 35}_{A}$ & {\bf 27} & {\bf 10} \\
\hline
1, $\frac{1}{2}$ & $-1$, $\frac{3}{2}$ & $\sqrt{\frac{10}{560}}$ & $-\sqrt{\frac{20}{105}}$
& $\sqrt{\frac{25}{108}}$ & $\sqrt{\frac{98}{432}}$ & $-\sqrt{\frac{8}{56}}$
& $\sqrt{\frac{8}{42}}$ \\
1, $\frac{1}{2}$ & $-1$, $\frac{1}{2}$ & $\sqrt{\frac{160}{560}}$ & $\sqrt{\frac{20}{105}}$
& $-\sqrt{\frac{4}{108}}$ & $\sqrt{\frac{128}{432}}$ & $\sqrt{\frac{8}{56}}$
& $\sqrt{\frac{2}{42}}$ \\
0, 1 & 0, 2 & $-\sqrt{\frac{3}{560}}$ & $\sqrt{\frac{6}{105}}$ & $-\sqrt{\frac{30}{108}}$
& $-\sqrt{\frac{15}{432}}$ &
$-\sqrt{\frac{15}{56}}$ & $\sqrt{\frac{15}{42}}$ \\
0, 1 & 0, 1 & $-\sqrt{\frac{45}{560}}$ & $-\sqrt{\frac{40}{105}}$ & $-\sqrt{\frac{32}{108}}$
& $\sqrt{\frac{25}{432}}$ &
$\sqrt{\frac{9}{56}}$ & $\sqrt{\frac{1}{42}}$ \\
0, 0 & 0, 1 & $\sqrt{\frac{270}{560}}$ & $-\sqrt{\frac{15}{105}}$ & $-\sqrt{\frac{12}{108}}$
& $-\sqrt{\frac{6}{432}}$ &
$-\sqrt{\frac{6}{56}}$ & $-\sqrt{\frac{6}{42}}$ \\
$-1$, $\frac{1}{2}$ & 1, $\frac{3}{2}$ & $-\sqrt{\frac{72}{560}}$ & $\sqrt{\frac{4}{105}}$
& $-\sqrt{\frac{5}{108}}$ & $\sqrt{\frac{160}{432}}$ & $-\sqrt{\frac{10}{56}}$
& $-\sqrt{\frac{10}{42}}$ \\
\hline
\end{tabular}
\hskip 1cm
\begin{tabular}{rrrr}
\hline
$Y=0$ & $I=0$ & {\bf 64} & {\bf 27} \\
\hline
1, $\frac{1}{2}$ & $-1$, $\frac{1}{2}$ & $-\sqrt{\frac{4}{7}}$ & $-\sqrt{\frac{3}{7}}$ \\
0, 1 & 0, 1 & $\sqrt{\frac{3}{7}}$ & $-\sqrt{\frac{4}{7}}$ \\
\hline
\end{tabular}
\vskip 0.7cm
\end{center}
\end{table*}
\newpage
\begin{table*}
\begin{center}
\begin{tabular}{rrrr}
\hline
$Y=-1$ & $I=\frac{5}{2}$ & {\bf 81} & {\bf 64} \\
\hline
0, 1 & $-1$, $\frac{3}{2}$ & $\sqrt{\frac{3}{5}}$ & $\sqrt{\frac{2}{5}}$ \\
$-1$, $\frac{1}{2}$ & 0, 2 & $\sqrt{\frac{2}{5}}$ & $-\sqrt{\frac{3}{5}}$ \\
\hline
\end{tabular}
\hskip 1cm
\begin{tabular}{rrrrrrrr}
\hline
$Y=-1$ & $I=\frac{3}{2}$ & {\bf 81} & {\bf 64} & ${\bf 35}_{S}$ & ${\bf 35}_{A}$
& {\bf 28} & {\bf 27} \\
\hline
1, $\frac{1}{2}$ & $-2$, 1 & $\sqrt{\frac{20}{160}}$ & $\sqrt{\frac{5}{35}}$
& $-\sqrt{\frac{36}{144}}$
& $\sqrt{\frac{36}{288}}$
& $-\sqrt{\frac{4}{40}}$ & $\sqrt{\frac{36}{140}}$ \\
0, 1 & $-1$, $\frac{3}{2}$ & $-\sqrt{\frac{1}{160}}$ & $-\sqrt{\frac{9}{35}}$
& $\sqrt{\frac{5}{144}}$
& $\sqrt{\frac{125}{288}}$ & $\sqrt{\frac{5}{40}}$ & $\sqrt{\frac{20}{140}}$\\
0, 1 & $-1$, $\frac{1}{2}$ & $\sqrt{\frac{40}{160}}$ & $\sqrt{\frac{10}{35}}$
& $\sqrt{\frac{32}{144}}$
& $\sqrt{\frac{8}{288}}$ & $\sqrt{\frac{8}{40}}$ & $-\sqrt{\frac{2}{140}}$\\
0, 0 & $-1$, $\frac{3}{2}$ & $\sqrt{\frac{45}{160}}$ & $-\sqrt{\frac{5}{35}}$
& $\sqrt{\frac{1}{144}}$
& $\sqrt{\frac{25}{288}}$ & $-\sqrt{\frac{9}{40}}$ & $-\sqrt{\frac{36}{140}}$\\
$-1$, $\frac{1}{2}$ & 0, 2 & $-\sqrt{\frac{9}{160}}$ & $\sqrt{\frac{1}{35}}$ &
$-\sqrt{\frac{45}{144}}$
& $\sqrt{\frac{45}{288}}$
& $\sqrt{\frac{5}{40}}$ & $-\sqrt{\frac{45}{140}}$ \\
$-1$, $\frac{1}{2}$ & 0, 1 & $\sqrt{\frac{45}{160}}$ & $-\sqrt{\frac{5}{35}}$
& $-\sqrt{\frac{25}{144}}$
& $-\sqrt{\frac{49}{288}}$
& $\sqrt{\frac{9}{40}}$ & $\sqrt{\frac{1}{140}}$ \\
\hline
\end{tabular}
\vskip 0.7cm
\begin{tabular}{rrrrrrrr}
\hline
$Y=-1$ & $I=\frac{1}{2}$ & {\bf 81} & {\bf 64} & ${\bf 35}_{S}$ & ${\bf 35}_{A}$
& {\bf 27} & {\bf 10} \\
\hline
1, $\frac{1}{2}$ & $-2$, 1 & $\sqrt{\frac{2}{70}}$ & $-\sqrt{\frac{4}{35}}$
& $\sqrt{\frac{18}{144}}$
& $\sqrt{\frac{18}{72}}$ &
$-\sqrt{\frac{18}{112}}$ & $\sqrt{\frac{18}{56}}$ \\
1, $\frac{1}{2}$ & $-2$, 0 & $\sqrt{\frac{9}{70}}$ & $\sqrt{\frac{8}{35}}$
& $-\sqrt{\frac{4}{144}}$
& $\sqrt{\frac{16}{72}}$ &
$\sqrt{\frac{36}{112}}$ & $\sqrt{\frac{4}{56}}$ \\
0, 1 & $-1$, $\frac{3}{2}$ & $-\sqrt{\frac{1}{70}}$ & $\sqrt{\frac{2}{35}}$
& $-\sqrt{\frac{64}{144}}$
& $-\sqrt{\frac{4}{72}}$ &
$-\sqrt{\frac{16}{112}}$ & $\sqrt{\frac{16}{56}}$ \\
0, 1 & $-1$, $\frac{1}{2}$ & $-\sqrt{\frac{4}{70}}$ & $-\sqrt{\frac{18}{35}}$
& $-\sqrt{\frac{25}{144}}$
& $\sqrt{\frac{1}{72}}$ &
$\sqrt{\frac{25}{112}}$ & $\sqrt{\frac{1}{56}}$ \\
0, 0 & $-1$, $\frac{1}{2}$ & $\sqrt{\frac{36}{70}}$ & $-\sqrt{\frac{2}{35}}$
& $-\sqrt{\frac{25}{144}}$
& $\sqrt{\frac{1}{72}}$ &
$-\sqrt{\frac{9}{112}}$ & $-\sqrt{\frac{9}{56}}$ \\
$-1$, $\frac{1}{2}$ & 0, 1 & $-\sqrt{\frac{18}{70}}$ & $\sqrt{\frac{1}{35}}$
& $-\sqrt{\frac{8}{144}}$
& $\sqrt{\frac{32}{72}}$ &
$-\sqrt{\frac{8}{112}}$ & $-\sqrt{\frac{8}{56}}$ \\
\hline
\end{tabular}
\hskip 1cm
\begin{tabular}{rrrr}
\hline
$Y=-2$ & $I=2$ & {\bf 81} & {\bf 64} \\
\hline
0, 1 & $-2$, 1 & $\sqrt{\frac{2}{5}}$ & $\sqrt{\frac{3}{5}}$ \\
$-1$, $\frac{1}{2}$ & $-1$, $\frac{3}{2}$ & $\sqrt{\frac{3}{5}}$ & $-\sqrt{\frac{2}{5}}$ \\
\hline
\end{tabular}
\vskip 0.7cm
\begin{tabular}{rrrrrrrr}
\hline
$Y=-2$ & $I=1$ & {\bf 81} & {\bf 64} & ${\bf 35}_{S}$ & ${\bf 35}_{A}$
& {\bf 28} & {\bf 27} \\
\hline
1, $\frac{1}{2}$ & $-3$, $\frac{1}{2}$ & $\sqrt{\frac{1}{20}}$ & $\sqrt{\frac{16}{105}}$
& $-\sqrt{\frac{6}{36}}$ & $\sqrt{\frac{3}{36}}$
& $-\sqrt{\frac{1}{30}}$ & $\sqrt{\frac{72}{140}}$ \\
0, 1 & $-2$, 1 & 0 & $-\sqrt{\frac{25}{105}}$ & $\sqrt{\frac{6}{36}}$
& $\sqrt{\frac{12}{36}}$
& $\sqrt{\frac{4}{30}}$ & $\sqrt{\frac{18}{140}}$ \\
0, 1 & $-2$, 0 & $\sqrt{\frac{3}{20}}$  & $\sqrt{\frac{48}{105}}$ & $\sqrt{\frac{8}{36}}$
& $\sqrt{\frac{1}{36}}$
& $\sqrt{\frac{3}{30}}$ & $-\sqrt{\frac{6}{140}}$ \\
0, 0 & $-2$, 1 & $\sqrt{\frac{6}{20}}$  & $-\sqrt{\frac{6}{105}}$ & $-\sqrt{\frac{1}{36}}$
& $\sqrt{\frac{8}{36}}$
& $-\sqrt{\frac{6}{30}}$ & $-\sqrt{\frac{27}{140}}$ \\
$-1$, $\frac{1}{2}$ & $-1$, $\frac{3}{2}$ & $-\sqrt{\frac{2}{20}}$ & $\sqrt{\frac{2}{105}}$
& $-\sqrt{\frac{12}{36}}$ & $\sqrt{\frac{6}{36}}$
& $\sqrt{\frac{8}{30}}$ & $-\sqrt{\frac{16}{140}}$ \\
$-1$, $\frac{1}{2}$ & $-1$, $\frac{1}{2}$ & $\sqrt{\frac{8}{20}}$ & $-\sqrt{\frac{8}{105}}$
& $-\sqrt{\frac{3}{36}}$ & $-\sqrt{\frac{6}{36}}$
& $\sqrt{\frac{8}{30}}$ & $\sqrt{\frac{1}{140}}$ \\
\hline
\end{tabular}
\hskip 1cm
\begin{tabular}{rrrrrr}
\hline
$Y=-2$ & $I=0$ & {\bf 81} & ${\bf 35}_{S}$ & ${\bf 35}_{A}$ & {\bf 10} \\
\hline
1, $\frac{1}{2}$ & $-3$, $\frac{1}{2}$ & $\sqrt{\frac{3}{56}}$ & $\sqrt{\frac{1}{36}}$
& $\sqrt{\frac{25}{72}}$ & $\sqrt{\frac{8}{14}}$ \\
0, 1 & $-2$, 1 & $-\sqrt{\frac{2}{56}}$ & $-\sqrt{\frac{24}{36}}$
& $-\sqrt{\frac{6}{72}}$ & $\sqrt{\frac{3}{14}}$ \\
0, 0 & $-2$, 0 & $\sqrt{\frac{27}{56}}$ & $-\sqrt{\frac{9}{36}}$
& $\sqrt{\frac{9}{72}}$ & $-\sqrt{\frac{2}{14}}$ \\
$-1$, $\frac{1}{2}$ & $-1$, $\frac{1}{2}$ & $-\sqrt{\frac{24}{56}}$
& $-\sqrt{\frac{2}{36}}$ & $\sqrt{\frac{32}{72}}$ &
$-\sqrt{\frac{1}{14}}$ \\
\hline
\end{tabular}
\vskip 0.7cm
\begin{tabular}{rrrr}
\hline
$Y=-3$ & $I=\frac{3}{2}$ & {\bf 81} & {\bf 64} \\
\hline
0, 1 & $-3$, $\frac{1}{2}$ & $\sqrt{\frac{1}{5}}$ & $\sqrt{\frac{4}{5}}$ \\
$-1$, $\frac{1}{2}$ & $-2$, 1 & $\sqrt{\frac{4}{5}}$ & $-\sqrt{\frac{1}{5}}$ \\
\hline
\end{tabular}
\hskip 1cm
\begin{tabular}{rrrrrr}
\hline
$Y=-3$ & $I=\frac{1}{2}$ & {\bf 81} & ${\bf 35}_{S}$ & ${\bf 35}_{A}$ & {\bf 28} \\
\hline
0, 1 & $-3$, $\frac{1}{2}$ & $\sqrt{\frac{1}{32}}$ & $\sqrt{\frac{81}{144}}$
& $\sqrt{\frac{81}{288}}$ & $\sqrt{\frac{1}{8}}$ \\
0, 0 & $-3$, $\frac{1}{2}$ & $\sqrt{\frac{9}{32}}$ & $-\sqrt{\frac{25}{144}}$
& $\sqrt{\frac{121}{288}}$ & $-\sqrt{\frac{1}{8}}$ \\
$-1$, $\frac{1}{2}$ & $-2$, 1 & $-\sqrt{\frac{4}{32}}$ & $-\sqrt{\frac{36}{144}}$
& $\sqrt{\frac{36}{288}}$ & $\sqrt{\frac{4}{8}}$ \\
$-1$, $\frac{1}{2}$ & $-2$, 0 & $\sqrt{\frac{18}{32}}$ & $-\sqrt{\frac{2}{144}}$
& $-\sqrt{\frac{50}{288}}$ & $\sqrt{\frac{2}{8}}$ \\
\hline
\end{tabular}
\vskip 0.7cm
\begin{tabular}{rrr}
\hline
$Y=-4$ & $I=1$ & {\bf 81} \\
\hline
$-1$, $\frac{1}{2}$ & $-3$, $\frac{1}{2}$ & 1 \\
\hline
\end{tabular}
\hskip 1cm
\begin{tabular}{rrr}
\hline
$Y=-4$ & $I=0$ & {\bf 28} \\
\hline
$-1$, $\frac{1}{2}$ & $-3$, $\frac{1}{2}$ & 1 \\
\hline
\end{tabular}
\vskip 0.7cm
\end{center}
\end{table*}

\begin{table*}
\caption{Wigner $D$ functions.} \vskip 0.7cm
\begin{center}
\begin{tabular}{ccccccccccccccccccc}
\hline $D_{ab}^{\lambda}$ & $p$ & $n$ & $\Lambda$ & $\Sigma^{+}$ &
$\Sigma^{0}$ & $\Sigma^{-}$ & $\Xi^{0}$ & $\Xi^{-}$ &
$\Delta^{++}$ & $\Delta^{+}$ & $\Delta^{0}$ & $\Delta^{-}$ &
$\Sigma^{*+}$ & $\Sigma^{*0}$ & $\Sigma^{*-}$ & $\Xi^{*0}$ &
$\Xi^{*-}$ &$\Omega^{-}$\\
\hline $D_{33}^{8}$ & $-\frac{7}{30}$ & $\frac{7}{30}$ & $0$ &
$-\frac{1}{6}$ & $0$ & $\frac{1}{6}$ & $\frac{1}{15}$ &
$-\frac{1}{15}$ & $-\frac{3}{8}$ & $-\frac{1}{8}$ & $\frac{1}{8}$
& $\frac{3}{8}$ & $-\frac{1}{4}$ &
$0$ & $\frac{1}{4}$ & $-\frac{1}{8}$ & $\frac{1}{8}$ & $0$\\
$D_{38}^{8}$ & $\frac{\sqrt{3}}{30}$ & $-\frac{\sqrt{3}}{30}$ &
$0$ & $\frac{\sqrt{3}}{6}$ & $0$ & $-\frac{\sqrt{3}}{6}$ &
$\frac{2\sqrt{3}}{15}$ & $-\frac{2\sqrt{3}}{15}$ &
$\frac{\sqrt{3}}{8}$ & $\frac{\sqrt{3}}{24}$ &
$-\frac{\sqrt{3}}{24}$ & $-\frac{\sqrt{3}}{8}$ &
$\frac{\sqrt{3}}{12}$ & $0$ & $-\frac{\sqrt{3}}{12}$
& $\frac{\sqrt{3}}{24}$ & $-\frac{\sqrt{3}}{24}$ & $0$ \\
$D_{83}^{8}$ & $-\frac{\sqrt{3}}{30}$ & $-\frac{\sqrt{3}}{30}$ &
$\frac{\sqrt{3}}{10}$ & $-\frac{\sqrt{3}}{10}$ &
$-\frac{\sqrt{3}}{10}$ & $-\frac{\sqrt{3}}{10}$ &
$\frac{2\sqrt{3}}{15}$ & $\frac{2\sqrt{3}}{15}$ &
$-\frac{\sqrt{3}}{8}$ & $-\frac{\sqrt{3}}{8}$ &
$-\frac{\sqrt{3}}{8}$ & $-\frac{\sqrt{3}}{8}$ & $0$ & $0$ &
$0$ & $\frac{\sqrt{3}}{8}$ & $\frac{\sqrt{3}}{8}$ & $\frac{\sqrt{3}}{4}$\\
$D_{88}^{8}$ & $\frac{3}{10}$ & $\frac{3}{10}$ & $\frac{1}{10}$
& $-\frac{1}{10}$ & $-\frac{1}{10}$ & $-\frac{1}{10}$ & $-\frac{1}{5}$
& $-\frac{1}{5}$ & $\frac{1}{8}$ & $\frac{1}{8}$ & $\frac{1}{8}$
& $\frac{1}{8}$ & $0$ & $0$ & $0$ & $-\frac{1}{8}$ & $-\frac{1}{8}$
&$-\frac{1}{4}$ \\
$D_{33}^{10}$ & $-\frac{1}{15}$ & $\frac{1}{15}$ & $0$ & $\frac{1}{15}$ & $0$
& $-\frac{1}{15}$ & $-\frac{1}{15}$ & $\frac{1}{15}$ & $0$ & $0$ & $0$
& $0$ & $0$ & $0$ & $0$ & $0$ & $0$ & $0$ \\
$D_{33}^{\bar{10}}$ & $-\frac{1}{15}$ & $\frac{1}{15}$ & $0$ &
$\frac{1}{15}$ & $0$ & $-\frac{1}{15}$ & $-\frac{1}{15}$ &
$\frac{1}{15}$ & $0$ & $0$ & $0$
& $0$ & $0$ & $0$ & $0$ & $0$ & $0$ & $0$\\
$D_{33}^{27}$ & $-\frac{4}{135}$ & $\frac{4}{135}$ & $0$ & $0$ &
$0$ & $0$ & $\frac{4}{135}$  & $-\frac{4}{135}$ & $-\frac{1}{21}$
& $-\frac{1}{63}$ & $\frac{1}{63}$ & $\frac{1}{21}$ &
$\frac{1}{21}$ &
$0$ & $-\frac{1}{21}$ & $\frac{4}{63}$ & $-\frac{4}{63}$ & $0$\\
$D_{83}^{27}$ & $-\frac{2\sqrt{3}}{135}$ &
$-\frac{2\sqrt{3}}{135}$ & $\frac{6\sqrt{3}}{135}$ &
$\frac{2\sqrt{3}}{405}$ & $\frac{2\sqrt{3}}{405}$ &
$\frac{2\sqrt{3}}{405}$ & $-\frac{2\sqrt{3}}{135}$ &
$-\frac{2\sqrt{3}}{135}$ & $-\frac{\sqrt{3}}{63}$ &
$-\frac{\sqrt{3}}{63}$ & $-\frac{\sqrt{3}}{63}$ &
$-\frac{\sqrt{3}}{63}$ & $\frac{5\sqrt{3}}{189}$ &
$\frac{5\sqrt{3}}{189}$ & $\frac{5\sqrt{3}}{189}$ &
$\frac{\sqrt{3}}{63}$
& $\frac{\sqrt{3}}{63}$ & $-\frac{\sqrt{3}}{21}$\\
\hline
\end{tabular}
\vskip 0.7cm
\begin{tabular}{cccccccccc}
\hline $D_{ab}^{\lambda}$ & $\Lambda\Sigma^{0}$ & $p\Delta^{+}$ &
$n\Delta^{0}$ & $\Lambda\Sigma^{*0}$ & $\Sigma^{+}\Sigma^{*+}$ &
$\Sigma^{0}\Sigma^{*0}$ & $\Sigma^{-}\Sigma^{*-}$ &
$\Xi^{0}\Xi^{*0}$ & $\Xi^{-}\Xi^{*-}$\\
\hline $D_{33}^{8}$ & $-\frac{\sqrt{3}}{10}$ &
$\frac{2\sqrt{5}}{15}$ & $\frac{2\sqrt{5}}{15}$ &
$\frac{\sqrt{15}}{15}$ & $-\frac{\sqrt{5}}{15}$ & $0$
& $\frac{\sqrt{5}}{15}$ & $-\frac{\sqrt{5}}{15}$ & $\frac{\sqrt{5}}{15}$\\
$D_{38}^{8}$ & $-\frac{1}{10}$ & $0$ & $0$ & $0$ & $0$ & $0$ & $0$
& $0$ & $0$\\
$D_{83}^{8}$ & $0$ & $0$ & $0$ & $0$ &
$-\frac{\sqrt{15}}{15}$ & $-\frac{\sqrt{15}}{15}$ &
$-\frac{\sqrt{15}}{15}$ & $-\frac{\sqrt{15}}{15}$ &
$-\frac{\sqrt{15}}{15}$\\
$D_{88}^{8}$ & $0$ & $0$ & $0$ & $0$ & $0$ & $0$ & $0$ & $0$ &
$0$\\
$D_{33}^{10}$ & $\frac{\sqrt{3}}{15}$ & $0$ & $0$ & $0$ & $0$ &
$0$ &
$0$ & $0$ & $0$\\
$D_{33}^{\bar{10}}$ & $-\frac{\sqrt{3}}{15}$ &
$\frac{\sqrt{5}}{15}$ & $\frac{\sqrt{5}}{15}$ & $0$ &
$\frac{\sqrt{5}}{15}$ & $0$
& $-\frac{\sqrt{5}}{15}$ & $\frac{\sqrt{5}}{15}$ & $-\frac{\sqrt{5}}{15}$\\
$D_{33}^{27}$ & $\frac{4\sqrt{3}}{135}$ & $\frac{\sqrt{5}}{270}$ &
$\frac{\sqrt{5}}{270}$ & $-\frac{\sqrt{15}}{135}$ &
$-\frac{\sqrt{5}}{90}$ & $0$
& $\frac{\sqrt{5}}{90}$ & $\frac{\sqrt{5}}{135}$ & $-\frac{\sqrt{5}}{135}$\\
$D_{83}^{27}$ & $0$ & $0$ & $0$ & $0$ & $-\frac{2\sqrt{15}}{405}$
& $-\frac{2\sqrt{15}}{405}$
& $-\frac{2\sqrt{15}}{405}$ & $\frac{\sqrt{15}}{135}$ & $\frac{\sqrt{15}}{135}$\\
\hline
\end{tabular}
\vskip 0.7cm
\end{center}
\end{table*}

\end{document}